\documentstyle[11pt,twoside,colloqOHP2010,epsfig]{article}
\begin{document}
\title{System Geometries and Transit$/$Eclipse Probabilities}
\author{K. von Braun$^1$, S. R. Kane$^1$, S. Mahadevan$^2$, G. Laughlin$^3$, A. Howard$^4$, \& D. R. Ciardi$^1$}
\affil{$^1$ NASA Exoplanet Science Institute, California Institute of Technology, MC 100-22, Pasadena, CA 91125, USA [kaspar@caltech.edu]} 
\affil{$^2$ Pennsylvania State University, 418 Davey Lab, University Park, PA 16802, USA} 
\affil{$^3$ Dept. of Astronomy \& Astrophysics, University of California at Santa Cruz, Santa Cruz, CA 95064, USA} 
\affil{$^4$ Astronomy Dept. \& Space Sciences Laboratory, University of California at Berkeley, Berkeley, CA 94720, USA} 
%
%
\begin{abstract}
Transiting exoplanets provide access to data to study the mass-radius relation and internal structure of extrasolar planets. Long-period transiting planets allow insight into planetary environments similar to the Solar System where, in contrast to hot Jupiters, planets are not constantly exposed to the intense radiation of their parent stars. Observations of secondary eclipses additionally permit studies of exoplanet temperatures and large-scale exo-atmospheric properties. We show how transit and eclipse probabilities are related to planet-star system geometries, particularly for long-period, eccentric orbits. The resulting target selection and observational strategies represent the principal ingredients of our photometric survey of known radial-velocity planets with the aim of detecting transit signatures (TERMS). 
%
\end{abstract}
%
%
\section{Transit/Eclipse Probabilities} \label{sec:probabilities}
\begin{figure}[ht]
\begin{center}
\plottwo{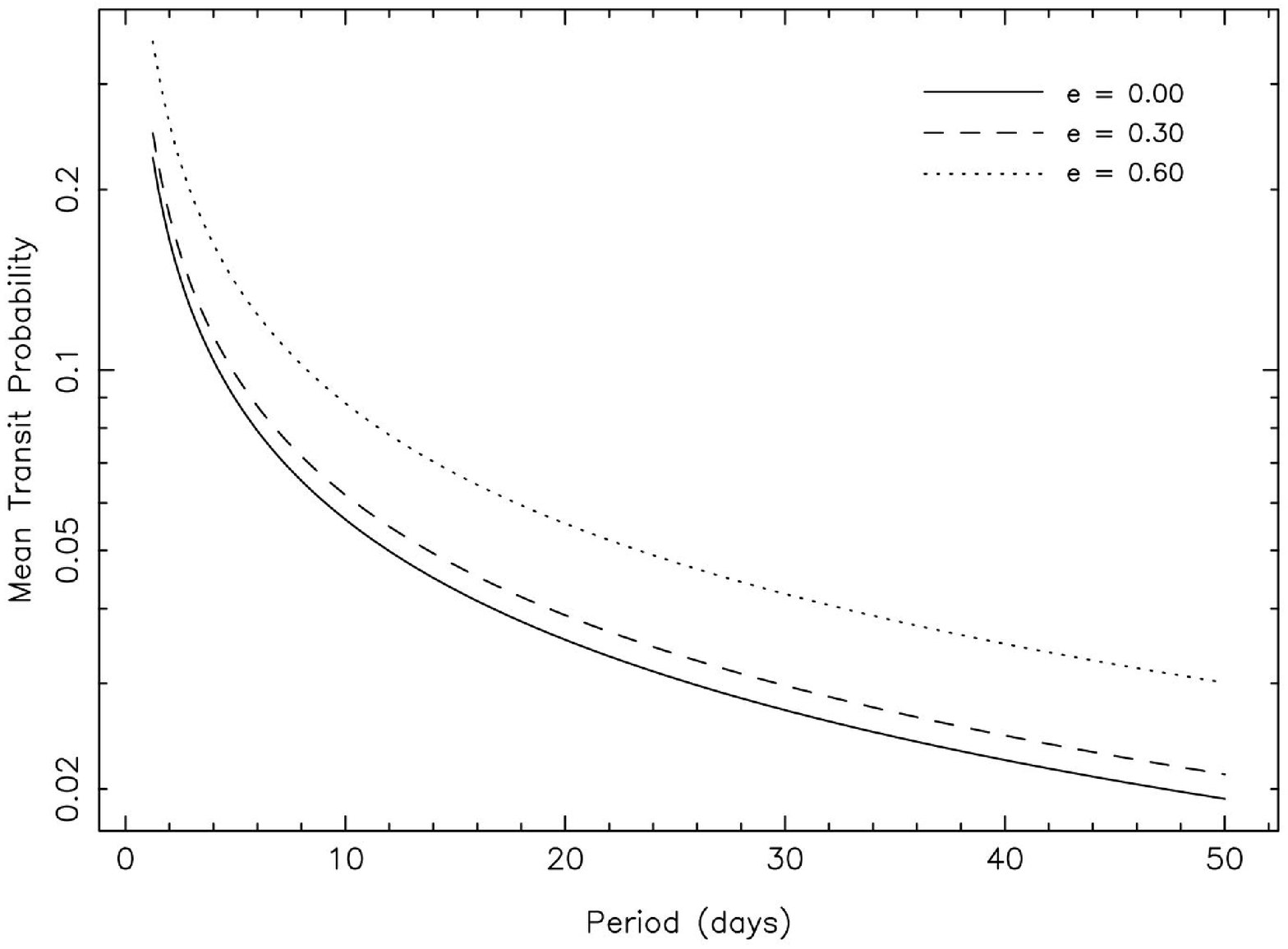}{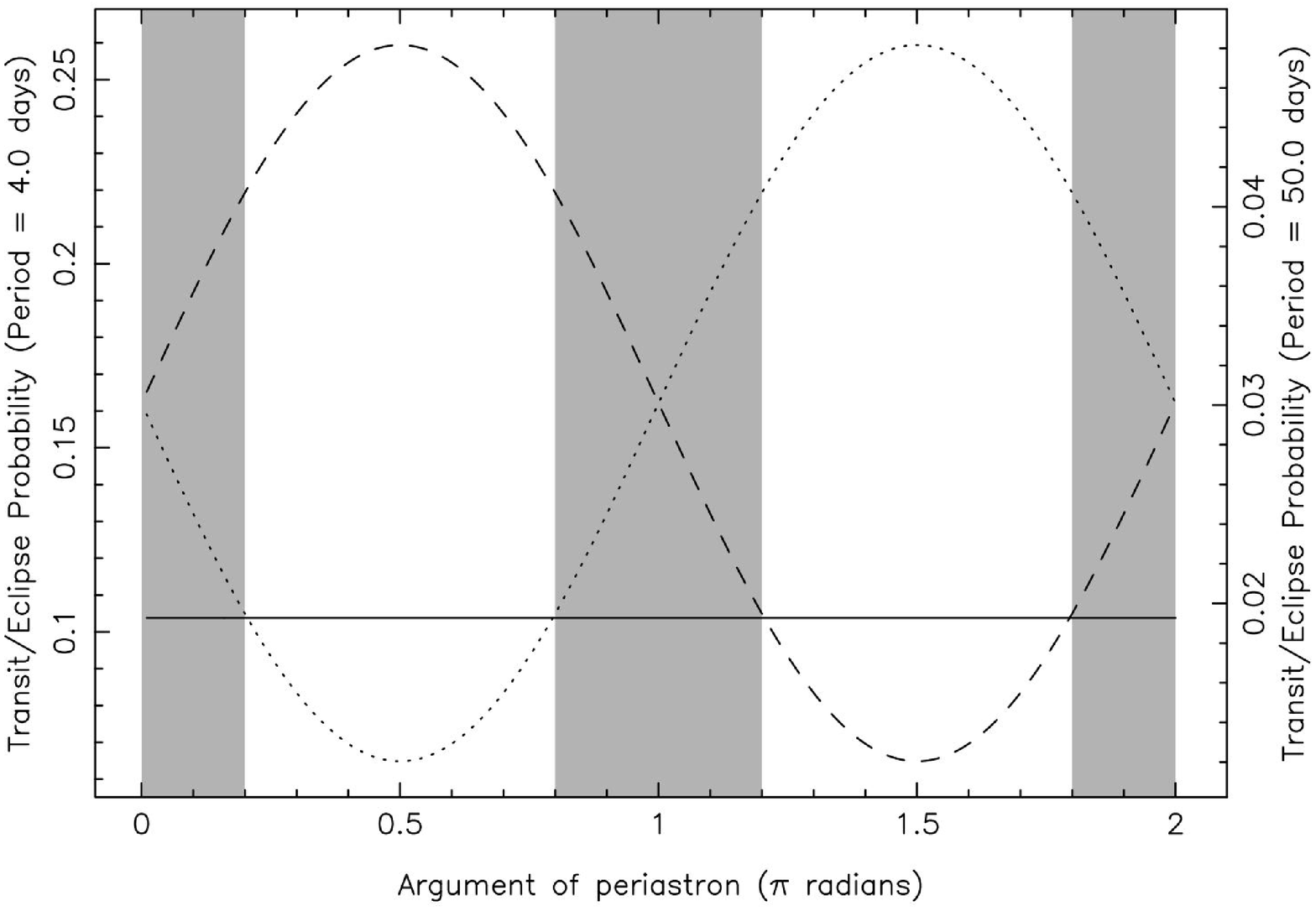}
\caption{{\bf Geometric transit and eclipse probabilities.} {\bf Left panel:} $P_t$ as a function of period for different eccentricities when averaged over all values of $\omega$. Solid line: circular orbit; dashed/dotted lines: $e$=0.3/0.6. Note how $P_t$ is higher for larger eccentricities. {\bf Right panel:} $P_t$ (dashed line) and $P_e$ (dotted line) as functions of $\omega$ for periods of 4.0 and 50.0 days (left and right ordinate, respectively). The shaded regions show ranges of $\omega$ in which both $P_t$ and $P_e$ exceed their respective counterpart for a circular orbit with the same period (solid line). We assume solar/Jupiter masses and radii for the star/planet.}
\label{fig:probabilities}
\end{center}
\end{figure}

The geometric probabilities with which an existing planet will transit its parent star ($P_t$) or be eclipsed by its parent star ($P_e$) are given by the following equations (Barnes 2007; Burke 2008; Kane \& von Braun 2008, 2009; von Braun \& Kane 2010)
\begin{equation}
  P_t = \frac{(R_p + R_\star)(1 + e \cos (\pi/2 - \omega))}{a (1 - e^2)}
  \label{eq:transit_prob}
\end{equation}
and
\begin{equation}
  P_e = \frac{(R_{p} + R_\star)(1 + e \cos (3\pi/2 - \omega))}{a (1 - e^2)},
  \label{eq:eclipse_prob}
\end{equation}
where $R_{p}$ and $R_\star$ are planetary and stellar radii, respectively, and $a$, $e$, and $\omega$ are the orbital semi-major axis, eccentricity, and argument of periastron. When averaging over all possible values of $\omega$, $P_t$ increases with increasing eccentricity (left panel in Fig. \ref{fig:probabilities}). For fixed eccentricities, equations \ref{eq:transit_prob} and \ref{eq:eclipse_prob} show that $P_e$ is highest for $\omega = 3\pi / 2$, whereas $P_t$ is highest for $\omega = \pi / 2$ (right panel in Fig. \ref{fig:probabilities}).

\section{TERMS}
The Transit Ephemerides Refinement and Monitoring Survey (TERMS) takes advantage of the combinations of (measured) $e$ and $w$ of  planets detected by the radial velocity (RV) method to search for transiting signatures among these planets (Kane et al. 2009). In the process, we use concurrently obtained RV data to minimize the size of the transit window for our photometric observations. 

\begin{figure}[ht]
\begin{center}
\epsfig{width=8cm,file=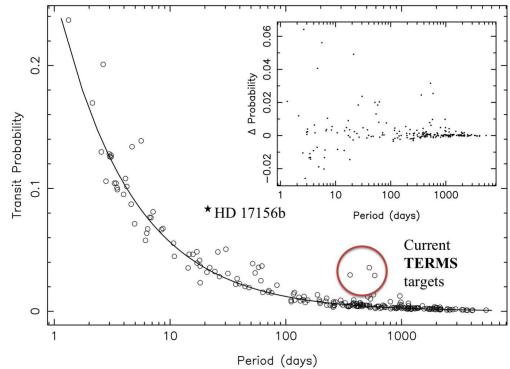}
\caption{{\bf TERMS:} $P_t$ for 203 exoplanets from the catalog of Butler et al. (2006), based on their tabulated orbital parameters (open circles). The solid line shows $P_t$ for circular orbits. Differences between $P_t$ for circular and actual orbits are shown in insert. The three current TERMS targets (due to their inflated $P_t$ values) are HD 156846b, HD 4113b, and HD20782b. For purposes of comparison, we assume solar/Jupiter masses and radii for the stars/planets.}
\label{fig:terms}
\end{center}
\end{figure}






\begin{references}
\reference Barnes, J. W. 2007, \pasp, 119, 986
\reference Burke, C. J. 2008, \apj, 679, 1566
\reference Butler, R. P., et al. 2006, \apj, 646, 505
\reference Kane, S. R., \& von Braun, K. 2008, \apj, 689, 492
\reference Kane, S. R., \& von Braun, K. 2009, \pasp, 121, 1096
\reference Kane, S. R., et al. 2009, \pasp, 121, 1386
\reference von Braun, K. \& Kane, S. R. 2010, ASP Conf. Series, 430, 551 
\end{references}
\end{document}